\documentclass[final,3p,times,twocolumn]{elsarticle}
\biboptions{sort&compress}
\usepackage{amsmath,amssymb}
\usepackage{graphicx}
\usepackage{hyperref}
\usepackage{amsmath,amssymb,amsfonts}
\usepackage{algorithmic}
\usepackage{amssymb}
\usepackage{algorithm}
\usepackage{graphicx}
\usepackage{textcomp}
\usepackage{xcolor}
\usepackage{times}
\usepackage{subcaption} 
\usepackage{stmaryrd}
\usepackage{balance}

\newtheorem{problem}{Problem}

\newtheorem{lemma}{Lemma}
\newtheorem{remark}{\bf Remark}
\newtheorem{assumption}{Assumption}
\newtheorem{theorem}{Theorem}

\newtheorem{proposition}{Proposition}

\newtheorem{proof}{Proof}

\DeclareMathDelimiter{\orbrack}{\mathopen}{operators}{"5D}{largesymbols}{"03}
\DeclareMathDelimiter{\clbrack}{\mathclose}{operators}{"5B}{largesymbols}{"02}
\def\intcc#1{\ensuremath{[#1]}}
\def\intoo#1{\ensuremath{\orbrack#1\clbrack}}
\def\intoc#1{\ensuremath{\orbrack#1]}}
\def\intco#1{\ensuremath{[#1\clbrack}}

\def\Hintcc#1{\ensuremath{\llbracket#1\rrbracket}}

\def\defas{\ensuremath{\mathrel{:=}}}

\DeclareMathOperator{\id}{id}

\def\Set#1#2{\ensuremath{
\left\{#1\,\middle|\,#2\right\}
}}

\def\norm#1{\|  #1 \| }

\def\abs#1{|  #1 | }

\def\BibTeX{{\rm B\kern-.05em{\sc i\kern-.025em b}\kern-.08em
    T\kern-.1667em\lower.7ex\hbox{E}\kern-.125emX}}

\journal{European Journal of Control}

\begin{document}

\begin{frontmatter}

\title{Learning Neural Network Safe Tracking Controllers \\ from Backward Reachable Sets}

\author[label1]{Yuezhu Xu}
\author[label2]{Mohamed Serry}
\author[label2]{Jun Liu}
\author[label1]{S. Sivaranjani\corref{cor1}}

%% Affiliations
\affiliation[label1]{
    organization={School of Industrial Engineering, Purdue University},
    addressline={315 N Grant Street}, 
    city={West Lafayette},
    postcode={47907}, 
    state={Indiana},
    country={USA}
}

\affiliation[label2]{
    organization={Department of Applied Mathematics, University of Waterloo},
    addressline={200 University Avenue West}, 
    city={Waterloo},
    postcode={N2L 3G1}, 
    state={Ontario},
    country={Canada}
}

%% Corresponding author info (optional)
\cortext[cor1]{Corresponding author. Email: sseetha@purdue.edu}

%% Footnotes / acknowledgments (merged into footnote text later)
\fntext[fn1]{The first two authors contributed equally to this manuscript.}
\fntext[fn2]{The work of Mohamed Serry and Jun Liu was supported in part by the Natural Sciences and Engineering Research Council of Canada, and in part by the Canada Research Chairs Program.}
\fntext[fn3]{The work of Yuezhu Xu and S. Sivaranjani was supported in part by the Air Force Office of Scientific Research grant FA9550-23-1-0492.}

\begin{abstract}
The design of tracking controllers that closely follow a reference trajectory while ensuring safety and robustness against disturbances is a challenging problem in the control of autonomous systems. 
In this work, we propose a neural network-based  safe tracking control framework for nonlinear discrete-time systems with  reach-avoid specifications in the presence of disturbances. Our approach begins with generation of a nominal trajectory using standard trajectory synthesis approaches, followed by construction of safe zonotopic backward reachable sets along the nominal trajectory. The states lying within the backward reachable sets are guaranteed to satisfy safe reachability specifications. Then, our key insight is to leverage the computed backward reachable sets to inform the architecture and training of a neural network-based tracking controller such that the neural network drives the system's states through these backward reachable sets, thereby improving the likelihood of safe reachability. We perform formal verification with conformal prediction to achieve statistical safety guarantees 
on the performance of the learned neural controller. The performance of our approach is illustrated through a numerical example on the discrete-time Dubin's car model.
\end{abstract}


\end{frontmatter}

\section{Introduction}
\label{sec:Introduction}
Control of autonomous systems (e.g., unmanned ground and aerial vehicles) frequently involves reach-avoid tasks, where the system's state needs to be driven safely to a target set, while satisfying  specific input and state constraints.  Such tasks have to be executed safely,  taking into account  the presence of model nonlinearity, uncertainties, and disturbances. Reachability control tasks are  typically handled by designing nominal trajectories, with associated nominal control inputs, and then  embedding tracking controllers  into the  control system to keep the system's states close to the prescribed nominal trajectories (see, e.g., \cite{GraichenTreuerZeitz07,LeeLeokMcClamroch10}). Designing tracking controllers that satisfy input and state constraints is a challenging problem, which has been extensively tackled in the literature, e.g., using control barrier functions \cite{FukudaSatohSakata20}, control contraction metrics \cite{LangsonChryssochoosRakovicMayne04,SinghLandryMajumdarSlotinePavone19,ZhaoLakshmananAckermanGahlawatPavoneHovakimyan22}, and backward reachable sets \cite{serry2024safe}. In general, existing methods often have computational (e.g., high memory requirement and CPU time) and/or theoretical  (e.g., restricted classes of systems and lack of feasibility guarantees) limitations, which motivates further research in this area.

Here, we consider the problem of designing  safe tracking controllers for reach-avoid problems in nonlinear discrete-time systems. Specifically, we will employ the notion of backward reachable sets (BRSs) for safe control design. Safe BRSs are  sets of states that are guaranteed to be driven to target sets, while satisfying input and state constraints. These sets have been shown to be effective in reach-avoid control tasks \cite{YangOzay21,YangZhangJeanninOzay22} and have in fact been used to synthesize safe optimization-based tracking controllers for these tasks under state/input constraints and disturbances \cite{serry2024safe}. While BRSs provide provable safety and reachability guarantees, the resulting optimization-based tracking controllers may have limited feasibility, as they can only be defined for states that can be driven into the BRSs. Further, such controllers are computationally complex and may have limited applicability especially in real-time applications where fast control evaluation and deployment are essential. 

In this paper, we consider the application of neural networks (NNs) for synthesis of safe  tracking controllers for this problem. We are particularly interested in NN-based controllers for two reasons: first, they are well-suited for deployment in online settings, since the computation of control actions involves simple operations once the NN is trained; second, NNs have strong generalization abilities \cite{arora2018optimization}, allowing for good performance not only on the training set but on unseen data as well, which is important in maintaining safety under disturbances. These advantages have made NNs a topic of extensive research interest within the control community, with applications to stability verification, stabilization and tracking controller synthesis \cite{abate2020formal, liu2024tool,xu2023learning, chen2021learning,chang2019neural, wu2023neural, hassan2022neural, wang2017neural, dai2021lyapunov, talukder2023robust}.

While training NN-based controllers is a well-studied problem for applications such as stabilization, extending these methods to meet reach-avoid specifications remains challenging due to additional requirements on transient behaviors to reach a target set while avoiding unsafe regions. Therefore, NN controllers trained without explicit guidance on safety and reachability may lead to unsafe or infeasible behaviors. Another key challenge, especially for control tasks, is the availability of `good' data points that can guide the training of NNs to satisfy the required state and input constraints, which are particularly challenging to obtain for complex reach-avoid type specifications.

Motivated by the utility of NNs and BRSs, we propose a new NN-based  safe tracking control framework for nonlinear discrete-time systems under disturbances, with training that is guided by BRSs. Our approach can be summarized as follows. We start with a nominal trajectory and associated nominal control input that can be synthesized using standard trajectory generation approaches (e.g., sampling-based and  optimization-based methods). We then compute zonotopic safe BRSs along the nominal trajectory that guarantee the satisfaction of safe reachability specifications. We then leverage these sets to guide the training of a NN-based tracking controller, which is designed to push the states through these BRSs, increasing the likelihood of safe tracking during reach-avoid maneuvers.  Leveraging already computed BRSs to inform NN training addresses the previously mentioned challenge of generating `good' training data for reach-avoid tasks with state/input constraints.  Once trained, our method yields a closed-form NN controller, thus overcoming the computational  limitations typically encountered in deploying optimization-based tracking controllers for online tasks.  Finally, we employ formal verification with conformal prediction to provide statistical safety guarantees on the tracking performance of the learned neural network controller.

In the literature, reachability analysis has been widely applied to NN control systems, with a primary emphasis on the formal verification of NN-based controller performance. For example, studies such as \cite{zhang2024reachability,everett2021reachability,everett2021efficient,zhang2023backward,xiang2019reachable,chen2023one,sharifi2024provable} explore reachability analysis for linear systems with neural controllers, or for NN-based models of dynamical systems, focusing largely on computational efficiency and reducing conservativeness. Some recent works have extended these methods to nonlinear systems with neural controllers \cite{zhang2024hybrid,siefert2025reachability}. However, despite substantial progress in developing reachability-based verification techniques for NN-controlled systems, their practical scalability to high-dimensional systems and long time horizons remains limited, primarily due to the significant nonlinearities introduced by NNs.

In contrast to verification, the use of reachability analysis during the training or synthesis of NN-based controllers is still relatively underexplored. For instance, \cite{koller2024set} proposes a set-based training method using zonotopic over-approximations to produce robust neural networks. In \cite{chung2021constrained}, a constrained NN training framework based on constrained zonotopes is introduced to ensure that NN outputs avoid predefined unsafe regions. The work in \cite{harapanahalli2024certified} presents a method for learning certifiable NN controllers, paired with a polytopic robust controlled-invariant set, using interval inclusions for systems described by ordinary differential equations. To the best of our knowledge, the current work is among the first to integrate NN-based learning with backward reachability analysis for tracking control in reach-avoid problems, offering a novel direction in the synthesis of safety-aware NN controllers.

The organization of this paper is as follows: the necessary preliminaries and notations are introduced in Section \ref{sec:Preliminaries}, the nonlinear system and the associated tracking problem, with the accompanying assumptions, are presented in Section \ref{sec:ProblemFormulation}, the proposed method is discussed in Section \ref{sec:ProposedMethod}, the  performance of the approach is illustrated through a numerical example in Section \ref{sec:NumericalExamples}, and future work are discussed in Section \ref{sec:Conclusion}.

\section{Preliminaries and Notations}
\label{sec:Preliminaries}
Let $\mathbb{R}$, $\mathbb{R}_+$,  $\mathbb{Z}$, and $\mathbb{Z}_{+}$ denote
the sets of real numbers, non-negative real numbers, integers, and
non-negative integers, respectively, and
$\mathbb{N} = \mathbb{Z}_{+} \setminus \{ 0 \}$.
Let $\intcc{a,b}$, $\intoo{a,b}$,
$\intco{a,b}$, and $\intoc{a,b}$
denote closed, open and half-open
intervals, respectively, with end points $a$ and $b$, and
 $\intcc{a;b}$, $\intoo{a;b}$,
$\intco{a;b}$, and $\intoc{a;b}$ stand for their discrete counterparts,
e.g.,~$\intcc{a;b} = \intcc{a,b} \cap \mathbb{Z}$, and
$\intco{1;4} = \{ 1,2,3 \}$.  In $\mathbb{R}^{n}
$, the relations $<$, $\leq$, $\geq$, and
$>$ are defined component-wise, e.g., $a < b$, where $a,b\in \mathbb{R}^{n}$, iff $a_i < b_i$ for
all $i\in  \intcc{1;n}$. For $a, b \in (\mathbb{R}\cup\{-\infty,\infty\})^ n$, with $a \leq b$,
the closed hyper-interval (or hyper-rectangle) $\Hintcc{a,b}$ denotes the set $\Set{x\in \mathbb{R}^{n}}{a\leq x\leq b}$. For $a\in\mathbb{R}$, we define $\lceil a\rceil$ to be the smallest integer no less than $a$. For $a,b\in \mathbb{R}^{n}$, $a.*b\in \mathbb{R}^{n}$ represents element-wise multiplication. 
The $n$-dimensional vectors  with entries  of zero and one are denoted by $0_{n}$ and $1_{n}$, respectively. For $x \in \mathbb{R}^{n}$, we define $\abs{x} = [\abs{x_1}~\ldots ~\abs{x_n}]^{\top}$. The $n\times n$ identity matrix is denoted $\id_{n}$. Given   $A\in \mathbb{R}^{n\times m}$,    $A^{\dagger}\in \mathbb{R}^{m\times n}$   denotes the Moore-Penrose inverse of $A$. 
The space of $n$-dimensional real vectors is equipped with the maximal norm $\norm{\cdot}_{\infty}$, $\ell_1$-norm  $\norm{\cdot}_1$, and the Euclidean norm $\norm{\cdot}_{2}$, and $\mathbb{B}_{n}$ denotes the $n$-dimensional closed unit ball w.r.t. $\norm{\cdot}_{\infty}$. For $n\times m$ matrices, $\norm{\cdot}_{\infty}$ corresponds to the matrix norm induced by the maximal norm.
Given $M, N \subseteq X$, $M\setminus N$ (set difference of $M$ and $N$) denotes the set  $\Set{x\in M}{x\not\in N}$, $M+N$ (Minkowski sum of $M$ and $N$) denotes the set 
$\Set{ y + z }{ y \in M, z \in N }$, and $M-N$ (Minkowski or Pontryagin difference of $M$ and $N$) denotes the set $\Set{z\in X}{z+N\subseteq M}$. The interior of a set $S\subseteq \mathbb{R}^{n}$ is denoted $\mathrm{int}(S)$.
    Given $c\in \mathbb{R}^{n}$ (center), and $G\in \mathbb{R}^{n\times m}$ (generators matrix), the zonotope associated with $c$ and $G$ is the set $c+G\mathbb{B}_{m}$ (see, e.g., \cite{Girard05}).
A standard $l$-layer feed-forward neural network (NN) structure $g$ is formulated recursively with the number of neurons in layer $\mathbf{L}_j$ being $d_j$, $j\in [1;l]$ and the input being $z=z^{(0)}\in \mathbb{R}^{d_0}$ as 
\begin{equation}\label{eq: standard NN}
\begin{aligned}
    &\mathbf{L}_j: \, z^{(i)} = \phi(v^{(i)})  \quad \forall i\in [1;l-1],\\
    &\mathbf{L}_l: \, 
    g(z)= z^{(l)}=v^{(l)}, \quad z^{(0)}=z,
\end{aligned}
\end{equation}
where $v^{(j)}=W_j z^{(j-1)}+b_j$, and $W_j\in \mathbb{R}^{d_{j}\times d_{j-1}}$ and $b_j\in\mathbb{R}^{d_{j}}$ represent the weight and bias for layer $\mathbf{L}_j$ respectively, and $\phi:\mathbb{R}^{d_j} \to \mathbb{R}^{d_j}$ is a nonlinear activation function that acts element-wise on its argument.

\section{Problem Formulation}
\label{sec:ProblemFormulation}
Consider a discrete nonlinear dynamical system:
\begin{equation}\label{eq:NonlinearSystem}
    x_{k+1}\in f(x_{k},u_{k})+\mathcal{W},~k\in \mathbb{Z}_{+},
\end{equation}
where $x_{k}\in\mathbb{R}^{n}$ and $u_{k}\in \mathbb{R}^{m}$ are the state and the input at step $k\in \mathbb{Z}_{+}$,  $f\colon \mathbb{R}^{n}\times \mathbb{R}^{m}\rightarrow \mathbb{R}^{n}$ represents the dynamics of  \eqref{eq:NonlinearSystem}, and $\mathcal{W}\subseteq \mathbb{R}^{n}$ is the disturbance set.

The disturbance set $\mathcal{W}\subseteq \mathbb{R}^{n}$ is  known and given by $\mathcal{W}=\Hintcc{-\overline{w},\overline{w}}$,~$\overline{w}\in \mathbb{R}_{+}^{n}$, $u_{k}\in \mathcal{U},~k\in \mathbb{Z}_{+}$, where $\mathcal{U}=\Hintcc{\underline{u},\overline{u}}$ is known, and the initial state $x_{0}$ belongs to a known set  $\mathcal{X}_{\mathrm{i}}=\Hintcc{\underline{x}_{\mathrm{i}},\overline{x}_{\mathrm{i}}}$.  Let $\mathcal{X}_{\mathrm{o}}=\Hintcc{\underline{x}_{\mathrm{o}},\overline{x}_{\mathrm{o}}}\subseteq \mathbb{R}^{n}$ be a hyper-rectangular operating domain and $\mathcal{X}_{\mathrm{u}}=\bigcup_{i=1}^{N_\mathrm{u}}\Hintcc{\underline{x}_{\mathrm{u}}^{(i)},\overline{x}_{\mathrm{u}}^{(i)}}\subseteq \mathbb{R}^{n}$ be  an unsafe set defined as a union of $N_\mathrm{u}$ hyper-rectangles.    We require that the system state is always inside the operating domain, while avoiding the unsafe set. Let  $\mathcal{X}_{\mathrm{t}}=\Hintcc{\underline{x}_{\mathrm{t}},\overline{x}_{\mathrm{t}}}\subseteq \mathbb{R}^{n}$ be a hyper-rectangular target set that we aim to drive the system's state into from the  initial set $\mathcal{X}_{\mathrm{i}}$. 
\begin{assumption}
 The function $f$ is twice continuously differentiable, and the matrix $D_{x}f(x,u)\in \mathbb{R}^{n\times n}$ is invertible for all $(x,u)\in \mathcal{X_{\mathrm{o}}}\times \mathcal{U}$, where $D_{x}f$  denotes the partial derivative of $f$ with respect to $x$. 
 \end{assumption}
  \begin{remark}
  The Jacobian of $f$ is invertible when $f$ represents the flow map of a continuous-time system. Furthermore, if $f$ is obtained via a numerical time-discretization scheme, such as the forward Euler method, its Jacobian is guaranteed to be invertible provided the time step is chosen sufficiently small. To illustrate this, consider
\[
f(x,u)=x+\Delta_t\, g(x,u),
\]
where $g:\mathbb{R}^n\times\mathbb{R}^m\to\mathbb{R}^n$ is the vector field of a continuous-time system and is twice continuously differentiable. Then
\begin{equation}
    D_x f(x,u) = \id_n + \Delta_t\, D_x g(x,u).
\end{equation}
With the condition 
\begin{equation}
    \Delta_t \;<\; \frac{1}{\displaystyle \sup_{(x,u)\in \mathcal{X}_0\times \mathcal{U}} 
\|D_x g(x,u)\|_\infty},
\end{equation}
we have that the matrix $D_x f(x,u)$ is invertible for all $(x,u)\in \mathcal{X}_0\times \mathcal{U}$. In this case, invertibility follows from the Neumann-series criterion 
$\| \Delta_t D_x g(x,u) \|_\infty < 1$, which ensures that $\id_n + \Delta_t D_x g(x,u)$ is nonsingular uniformly over the domain.
 \end{remark}

\begin{assumption} \label{assm: XiXt}
$\mathcal{X}_{\mathrm{i}}\subseteq \mathcal{X}_{\mathrm{o}}\setminus \mathcal{X}_{\mathrm{u}}$ and $\mathcal{X}_{\mathrm{t}}\subseteq \mathcal{X}_{\mathrm{o}}\setminus \mathcal{X}_{\mathrm{u}}$. 
\end{assumption}

  Let $N\in \mathbb{N}$ and $\intcc{0;N}$ be a finite-time horizon over which the reach-avoid specifications above  are to be satisfied for system \eqref{eq:NonlinearSystem}, assuming zero disturbance. Specifically, let $\{\tilde{x}_{k}\}_{k=0}^{N}$ and $\{\tilde{u}_{k}\}_{k=0}^{N-1}$ be given nominal state and input sequences. Note that such nominal state and input sequences can be constructed by means of   standard trajectory generation approaches such as  RRT  \cite{LaValleKuffnerJr01}  or trajectory optimization \cite{KurtzLin20,ChenZhanTomizuka19,ZhaoLinTomizuka18,SasfiZeilingerKohler23}. The nominal state and input sequences are designed to satisfy:
\begin{align*}
\tilde{x}_{k+1}&=f(\tilde{x}_{k},\tilde{u}_{k}),~k\in \intcc{0;N-1}\\
\tilde{u}_{k}&\in \mathcal{U},~k\in \intcc{0;N-1},\\
\tilde{x}_{k} &\in \mathrm{int}(\mathcal{X}_{\mathrm{o}})\setminus\mathcal{X}_{\mathrm{u}},~k\in \intcc{0;N},\\
\tilde{x}_{0}&\in \mathcal{X}_{\mathrm{i}},~
\tilde{x}_{N}\in \mathrm{int}(\mathcal{X}_{\mathrm{t}}).
\end{align*}
We state the safe tracking control design problem here.
\begin{problem}\label{prob:main}
Given the  nominal sequences $\{\tilde{x}_{k}\}_{k=0}^{N}$ and $\{\tilde{u}_{k}\}_{k=0}^{N-1}$, find  a NN  control law { $\mu:\mathbb{R}^{n}\times \intcc{0;N-1}\rightarrow \mathcal{U}$}  that drives the states of   system \eqref{eq:NonlinearSystem}  from the initial set  $\mathcal{X}_{\mathrm{i}}$  to the target set $\mathcal{X}_{\mathrm{t}}$ in $N$ steps, while staying in the operating domain $\mathcal{X}_{\mathrm{o}}$ and avoiding the unsafe set  $\mathcal{X}_{\mathrm{u}}$.
 \end{problem}

\section{Neural-Network Tracking Control using Backward Reachable Sets}\label{sec:ProposedMethod}
We now present our NN-based tracking control framework, which consists of first constructing backward reachable sets (BRSs) and then using them to guide the training of  NN controllers.

\subsection{Backward Reachable Sets} \label{sec:bakcward}
In this section, we  briefly outline the computations of BRSs, following the approach in \cite{serry2024safe}. We first  construct an initial  state tube that consists of the hyper-rectangles:
\begin{align*}
\mathcal{T}_{x,k}&=\tilde{x}_{k}+\Hintcc{-R_{x,k},R_{x,k}},~R_{x,k} \in \mathbb{R}_{+}^{n},~k\in \intcc{0;N},
\end{align*}
satisfying:
\begin{align*}
\mathcal{T}_{x,k}& \subseteq \mathcal{X}_{\mathrm{o}}\setminus \mathcal{X}_{\mathrm{u}},~k\in \intcc{0;N-1},\\ \mathcal{T}_{x,N}&\subseteq \mathcal{X}_{\mathrm{t}}.
\end{align*} 
We also construct an initial input tube consisting of the hyper-rectangles:
$$
\mathcal{T}_{u,k}=\tilde{u}_{k}+\Hintcc{-R_{u,k},R_{u,k}}\subseteq \mathcal{U},~R_{u,k}\in \mathbb{R}_{+}^{m},~k\in \intcc{0;N-1}.
$$

These tubes ensure satisfaction of the state and input constraints and  are also used to quantify linearization errors. Next, the nonlinear dynamics \eqref{eq:NonlinearSystem} is conservatively linearized along the nominal states and inputs as follows. We define, for $k\in \intcc{0;N-1}$, 
\begin{align*}
A_{k}&\defas D_{x}f(\tilde{x}_{k},\tilde{u}_{k}),\\
B_{k}&\defas D_{u}f(\tilde{x}_{k},\tilde{u}_{k}),\\
c_{k}&\defas f(\tilde{x}_{k},\tilde{u}_{k})-A_{k}\tilde{x}_{k}-B_{k}\tilde{u}_{k},
\end{align*}
where $D_{u}f$ denotes the partial derivative of $f$ with respect to  $u$. Note that $A_{k}$ is invertible for all $k\in \intcc{0;N-1}$ due to the invertibility assumption on $D_{x}f$ in Section \ref{sec:ProblemFormulation}.

For a fixed $k\in \intcc{0;N-1}$ and state and input hyper-rectangles $T_{x,k}=\tilde{x}_{k}+\Hintcc{-{r}_{x},{r}_{x}},r_{x}\in \mathbb{R}^{n}_{+}$, and $T_{u,k}=\tilde{u}_{k}+\Hintcc{-{r}_{u},{r}_{u}},~r_{u}\in \mathbb{R}^{m}_{+}$, we define the set $\mathsf{W}({T}_{x,k},{T}_{x,k})\subseteq \mathbb{R}^{n}$ such that it  over-estimates  the linearization errors over $T_{x,k}$ and $T_{u,k}$ and accounts  for the disturbance set $\mathcal{W}$. Particularly,
$$
 \mathsf{W}({T}_{x,k},{T}_{u,k})\defas  \mathcal{W}+\Hintcc{-\mathbf{e}({T}_{x,k},{T}_{u,k}),\mathbf{e}({T}_{x,k},{T}_{u,k})},
 $$
where
$$
(\mathbf{e}({T}_{x,k},{T}_{u,k}))_{i}\defas\frac{1}{2} [{r}_{x}^{\top}, {r}_{u}^{\top}]\bar{\mathrm{H}}_{f_{i}}{({T}_{x,k},{T}_{u,k})}[{r}_{x}; {r}_{u}],
$$
$i\in \intcc{1;n}$. Herein,   $\bar{\mathrm{H}}_{f_{i}}{({T}_{x,k},{T}_{u,k})}\in \mathbb{R}^{(n+m)\times(n+m)}$ 
is a matrix that satisfies:
$$
(\bar{\mathrm{H}}_{f_{i}}{({T}_{x,k},{T}_{u,k})})_{p,q}\geq\sup_{{T}_{x,k}\times{T}_{u,k}}\abs{\{\mathrm{Hess}[\mathcal{F}_{i}](\cdot)\}_{p,q}},
$$
$p,q\in \intcc{1;n+m}$, where $\mathcal{F}\colon \mathbb{R}^{n+m}\rightarrow \mathbb{R}^{n}$ is defined as $\mathcal{F}([x^\top,u^\top]^{\top})=f(x,u)$, $(x,u)\in \mathbb{R}^{n}\times \mathbb{R}^{m}$, and $\mathrm{Hess}[\mathcal{F}_{i}](z)$ denotes the Hessian of $\mathcal{F}_{i}$ evaluated at $z$ (recall the twice-differentiability assumption on $f$). 

Incorporating $\mathsf{W}$ ensures over-approximation of the nonlinear dynamics \eqref{eq:NonlinearSystem} using the linearized dynamics as highlighted in the following lemma.

\begin{lemma}\label{lem:ConservativeLinearization}
 For all $(x,u)\in {T}_{x,k}\times {T}_{u,k}$,
$$
f(x,u)+\mathcal{W}\subseteq  A_{k}x+B_{k}u+c_{k}+\mathsf{W}({T}_{x,k},{T}_{u,k}).
$$
\end{lemma}

With the linearized dynamics, we compute under-approximate zonotopic BRSs and simultaneously optimize the state and input tubes in order to maximize the sizes of the resulting BRSs. This yields  zonotopic safe BRSs,  $\Lambda_{k},~k\in \intcc{0;N}$,  of the form: 
$$
\Lambda_{k}=\tilde{x}_{k}+G^{k}\mathbb{B}_{q_{k}},~G^{k}\in \mathbb{R}^{n\times q_{k}},~q_{k}\in \mathbb{N},~k\in \intcc{0;N},
$$
an \textit{optimized} input tube consisting of  hyper-rectangles
$$\mathsf{T}_{u,k}=\tilde{u}_{k}+\Hintcc{-\mathsf{r}_{u,k},\mathsf{r}_{u,k}}\subseteq  \mathcal{T}_{u,k}\subseteq\mathcal{U},~\mathsf{r}_{u,k}\in \mathbb{R}_{+}^{m},
$$
$k\in \intcc{0;N-1}$, and an \textit{optimized }state tube consisting of  hyper-rectangles
$$\mathsf{T}_{x,k}=\tilde{x}_{k}+\Hintcc{-\mathsf{r}_{x,k},\mathsf{r}_{x,k}}\subseteq  \mathcal{T}_{x,k},~\mathsf{r}_{x,k}\in \mathbb{R}_{+}^{n},
$$
$~k\in \intcc{0;N-1}$. The BRSs $\Lambda_{i},~i\in \intcc{0;N}$ are defined such that:
\begin{align*}
\Lambda_{N}&=\mathsf{T}_{x,N},\\
\Lambda_{k}&\subseteq A_{k}^{-1}(\Psi_{k}+(-(c_{k}+B_{k}\mathsf{T}_{u,k})))\cap \mathsf{T}_{x,k},
\end{align*}
$k\in \intcc{0;N-1}$, where $\Psi_{k}$ satisfies:
\begin{align*}
\Psi_{k}&\subseteq (\Lambda_{k+1}-\mathsf{W}(\mathsf{T}_{x,k},\mathsf{T}_{u,k})).
\end{align*}
Each set $\Lambda_{k}$ under-approximates the  safe BRS (located in the safe hyper-rectangle $\mathsf{T}_{x,k}$), starting from $\Lambda_{k+1}$ under the conservatively linearized dynamics with input values in $\mathsf{T}_{u,k}$. 

To account for the effect of the disturbances represented by  set  $\mathcal{W}$, the BRSs are accompanied by their deflated versions:
$$
\tilde{\Lambda}_{k}=\tilde{x}_{k}+\tilde{G}^{k}\mathbb{B}_{q_{k}}\subseteq \Lambda_{k}-\mathcal{W},~\tilde{G}^{k}\in \mathbb{R}^{n\times q_{k}},~k\in \intcc{0;N},
$$
The sets $\tilde{\Lambda}_{k}$ and $\Psi_{k}$ are acquired by under-approximating Minkowski differences, with employment of the method in \cite[Lemma~7]{serry2024safe}. The construction of the sequence of BRSs  $\{\Lambda_{k}\}_{k=0}^{N}$, the deflated version $\{\tilde{\Lambda}_{k}\}_{k=0}^{N}$,  the state and  input sequences $\{\mathsf{T}_{x,k}\}_{k=0}^{N}$ and  $\{\mathsf{T}_{u,k}\}_{k=0}^{N-1}$ provides the following safety and reachability guarantees. 
\begin{theorem}\label{thm:brs}
For all $k\in \intcc{0;N-1}$,
$
\Lambda_{k}\subseteq \mathcal{X}_{\mathrm{o}}\setminus \mathcal{X}_{\mathrm{u}},
$
$
x\in \Lambda_{k}\Rightarrow \exists u\in \mathsf{T}_{u,k}\subseteq \mathcal{U}~\text{s.t.}~f(x,u)\in \tilde{\Lambda}_{k+1}\Rightarrow f(x,u)+\mathcal{W}\subseteq \Lambda_{k+1},
$
and 
$
\Lambda_{N}\subseteq \mathcal{X}_{\mathrm{t}}.
$
\end{theorem}
    The proofs of Lemma \ref{lem:ConservativeLinearization} and Theorem \ref{thm:brs} follow along the lines of \cite{serry2024safe} and are omitted for brevity.

\subsection{Architecture Design and Training of \\Neural Tracking Controllers}\label{sec:NN design and training}
With the BRSs, namely $\Lambda_k$, $k \in [0;N]$ computed\footnote{ In the computation of the BRSs, we consider a slightly inflated version of the disturbance set $\mathcal{W}$ to provide a margin that increases the safe tracking capability of the learned NN controller.}, we aim to obtain a NN tracking controller $\mu:\mathbb{R}^{n}\times \intcc{0;N-1}\rightarrow \mathcal{U}$ that drives the states in $\Lambda_k$ towards the center of $\tilde{\Lambda}_{k+1}$, that is, $\tilde{x}_{k+1}$,  $k \in [0;N-1]$. For each step $k$, $\mu(\cdot,k)$ is  designed to have  a NN architecture with feed-forward and multiplication structure, with parameters learned through optimization. 

\subsubsection{Neural Network Architecture Design}
We first present the architecture of the NN that will be utilized for tracking control. The NN structure is  designed to  (i) satisfy the input constraint $\mu(\cdot,k)\subseteq \mathcal{U},~k\in \intcc{0;N-1}$, and (ii) to ensure that the  control values generated by the NN match the the nominal control values when the states coincide with their nominal values $\{\tilde{x}_{k}\}_{k=0}^{N-1}$, that is, $\mu(\tilde{x}_{k},k)=\tilde{u}_{k},~k\in \intcc{0;N-1}$.  To guarantee these properties, the architecture of $\mu$ is designed as follows. Consider a time step $k\in \intcc{0;N-1}$. Let $\tilde{\mu}(\cdot,k)\colon \mathbb{R}^{n}\rightarrow \mathbb{R}^{m}$ be an $l$-layer feed-forward NN whose input is the state of the system \eqref{eq:NonlinearSystem} at time $k$, denoted by $x_k$. Then, we have $d_{0}=n$ and $d_{l}=m$. Define $x_k^{(0)}=x_k-\tilde{x}_k$. We design the first $l-2$ layers similar to the standard structure described  in \eqref{eq: standard NN} as
\begin{equation}
    \mathbf{L}_{k,j}: \, x_k^{(j)} = \phi(v_k^{(j)})  \quad \forall j\in [1;l-2], 
\end{equation}
with $v_k^{(j)}=W_{k,j} x_k^{(j-1)}+b_{k,j}$. The parameters $W_{k,j}$ and $b_{k,j}$ represent the weight and bias for layer $\mathbf{L}_{k,j}$, respectively, which are to be trained. For the $(l-1)^{\text{th}}$ layer, we set $d_{l-1}=d_0=n$ and first introduce a multiplication operation linking the normalized input and $\mathbf{L}_{k,l-1}$ as 
\begin{equation}
    x_k^{(l-1)}=x_k^{(0)}.*(W_{k,l-1} x_k^{(l-2)}+b_{k,l-1}).
\end{equation}
Then, we define 
\begin{equation}
    \mathbf{L}_{k,l}: x_k^{(l)}=\Phi(W_{k,l}x_k^{(l-1)})
\end{equation}
with $\Phi$ being the \textit{tanh} function and $W_{k,l}\in\mathbb{R}^{m\times n}$. 
Finally, the output is translated and scaled using a factor $R_{k,ul}\in \mathbb{R}^m$ 
\begin{equation} \label{eq:NN design last layer} 
    \tilde{\mu}(x_k,k)= R_{k,ul}~.*x_k^{(l)}+\tilde{u}_k.  
\end{equation}
To retain the learning ability of the NN, the scaling factor $R_{k,ul}$ is set as a trainable parameter. With the above architecture, $\tilde{\mu}(\cdot,k)$ satisfies the following properties:
\begin{lemma} \label{lem:NN structure}
We have   $\tilde{\mu}(\tilde{x}_k,k)=\tilde{u}_k$ and $\tilde{\mu}(x_{k},k)\in \Hintcc{\tilde{u}_k-|R_{k,ul}|, \tilde{u}_k+|R_{k,ul}|}$ .
\end{lemma}
\begin{proof}
If  $x_k=\tilde{x}_k$, then $x_k^{(0)}=x_k-\tilde{x}_k=0_{n}$.
Correspondingly, $x_k^{(l-1)}=x_k^{(0)}.*(W_{k,l-1} x_k^{(l-2)}+b_{k,l-1})=0_{n}$ and $x_k^{(l)}=\tanh(0_{m})=0_m$. By \eqref{eq:NN design last layer}, we have $\tilde{\mu}(\tilde{x}_k,k)=\tilde{u}_k$.
Also, since we use \textit{tanh} as the activation function $\Phi$ in layer $\mathbf{L}_{k,l-1}$, limiting $x_k^{(l-1)}$ in the range $\Hintcc{-1_{m},1_{m}}$, we have $\tilde{\mu}(x_k,k)\in\Hintcc{\tilde{u}_k-|R_{k,ul}|, \tilde{u}_k+|R_{k,ul}|}$.
\end{proof}
Finally,  to satisfy the input constraints, the control value $\mu(x_{k},k)$ is obtained by trimming $\tilde{\mu}(x_{k},k)$ as follows:
$$
(\mu(x_{k},k))_{j}=\begin{cases}
    (\tilde{\mu}(x_{k},k))_{j},& (\tilde{\mu}(x_{k},k))_{j}\in \intcc{\underline{u}_{j},\overline{u}_{j}},\\ \underline{u}_{j}, &(\tilde{\mu}(x_{k},k))_{j}<\underline{u}_{j},\\
    \overline{u}_{j},& (\tilde{\mu}(x_{k},k))_{j}>\overline{u}_{j},\ \ 
\end{cases}
$$
$ j\in \intcc{1;m}$. Note that since we employ the same network structure for all $k\in \intcc{0;N-1}$, we omit the index $k$ in the dimension $d_i$, $i\in \intcc{0;l}$ for brevity.
We also note that the NN training is designed to optimize the performance of the intermediate controller, $\tilde{\mu}$, rather than that of the actual controller, $\mu$. Empirical evidence indicates that, in the majority of cases, $\tilde{\mu}(x_k, k) \in \mathcal{U}$, ensuring that $\mu(x_k, k) = \tilde{\mu}(x_k, k)$. Consequently, additional clipping, as defined above, is rarely required and is only applied in exceptional scenarios where the inputs fall slightly outside the allowed range.

\subsubsection{Sampling Methods for Training Data}
Next, we discuss how we sample the states inside $\Lambda_k=\tilde{x}_k+G^k\mathbb{B}_{q_k}$ for time step $k\in[0;N-1]$, to construct the training data set.
Intuitively, the states near the boundary of $\Lambda_k$ are more deviated from the nominal trajectory, making them harder to drive towards $\tilde{x}_{k+1}$, $k\in[0;N-1]$. To address this challenge during the learning process, we use both extreme sampling (points near boundary) and uniform sampling to construct the training sets $\mathbf{S}_k$ as follows. For uniform sampling, we uniformly generate $H_k^u$ points $\{r_i\}_{i=1}^{H_k^u}$ randomly from $\mathbb{B}_{q_k}$. The uniform sampling set is then defined as $\mathbf{S}_k^u\triangleq\{\tilde{x}_k+G^kr_i\}_{i=1}^{H_k^u}$. For extreme sampling, we  generate $H_k^{ex}$ data points to form set $\mathbf{S}_k^{ex}$. The sampling data set $\mathbf{S}_k$ is then $\mathbf{S}_k=\mathbf{S}_k^u\bigcup\mathbf{S}_k^{ex}$. We randomly split $\mathbf{S}_k$ into training set $\mathbf{S}_{k, train}$ and validation set $\mathbf{S}_{k, val}$. We denote the training points at time step $k$ as $z_{k,h}$ for $h \in \intcc{1; H_k^{train}}$, and the validation points at time step $k$ as $z_{k,h}$ for $h \in \intcc{1;H_k^{val}}$.

\subsubsection{Optimization for Neural Network Training}

For each $k\in \intcc{0;N-1}$, the problem of training the NN tracking controller $\tilde{\mu}$ can be posed as the following optimization problem: 
\begin{equation} \label{opt}
    \begin{aligned}
    \text{minimize} \quad & P_k=\alpha_{k,1} P_{k,1}+\alpha_{k,2} P_{k,2}, \\
\end{aligned}
\end{equation}
where we have two penalty terms $P_{k,1}$ and $P_{k,2}$ in the objective function, defined as
\begin{equation} \label{eq:loss P1}
\begin{aligned}
    P_{k,1}=&
    \frac{1}{H_k^{train}}\sum\limits_{z_{k,h} \in \mathbf{S}_{k, train}}\Bigg[ \|d_{k,G}(z_{k,h})\|_{\infty} \\
    &+\lambda\exp\left(\max( \|d_{k,G}(z_{k,h})\|_{\infty} -1,0)\right)\Bigg], 
\end{aligned}
\end{equation}
with $d_{k,G}(\cdot)\triangleq(\tilde{G}^{k+1})^{\dagger}(f(\cdot,\tilde{\mu}(\cdot,k))-\tilde{x}_{k+1})$, and 
\begin{equation} \label{eq:loss P2}
    P_{k,2}=\|R_{k,ul}\|_1,
\end{equation}
with $\lambda>0$, $\alpha_{k,1}>0, \, \alpha_{k,2}>0$ being the respective weighting factors. Broadly speaking, the first term $P_{k,1}$ guides the state toward
the consecutive backward reachable set, and the second term $P_{k,2}$ is concerned with the input being inside the allowed range. We examine the construction of each term below.

\textbf{Penalty Term} ${P_{k,1}}:$ Note that for a given $k\in \intcc{0;N-1}$ and  state $x_k\in \Lambda_{k}$, if the  control law $\tilde{\mu}$ 
ensures that  
$\|d_{k,G}(x_k)\|_{\infty}\leq 1$,
then  $x_k$ is guaranteed to be driven to  $\Lambda_{k+1}$ under the closed-loop dynamics of \eqref{eq:NonlinearSystem}, where the smaller the value of $\|d_{k,G}(x_k)\|_{\infty}$, the closer the next state is to $\tilde{x}_{k+1}$ (the center of $\Lambda_{k+1}$). {To see this, let $\tilde{b}=(\tilde{G}^{k+1})^{\dagger}(f(x_k,\tilde{\mu}(x_k,k))-\tilde{x}_{k+1})\in \mathbb{B}_{q_{k+1}}$; then $f(x_k,\tilde{\mu}(x_k,k))=\tilde{x}_{k+1}+\tilde{G}^{k+1}\tilde{b}\in \tilde{\Lambda}_{k+1}$, and hence $f(x_k,\tilde{\mu}(x_k,k))+\mathcal{W}\subseteq \Lambda_{k+1}$.}
Based on this fact, we penalize the norm itself and additionally include an exponential penalty if the norm exceeds 1. With training points $z_{k,h}$, $h\in \intcc{1;H_k^{train}}$, which are sampled from $\Lambda_{k}$, we then have $P_{k,1}$ as  in \eqref{eq:loss P1}.

\textbf{Penalty Term} $P_{k,2}:$ According to Lemma \ref{lem:NN structure},  $\tilde{\mu}(x_k,k)\in\Hintcc{\tilde{u}_k-|R_{k,ul}|, \tilde{u}_k+|R_{k,ul}|}$.
Effectively, $\|R_{k,ul}\|_1$  represents the deviation of control value generated by the NN from the nominal input $\tilde{u}_k$, which we penalize by including term $P_{k,2}$ as defined in \eqref{eq:loss P2}.

\begin{remark}
The learned neural network controller $ \mu $ does not possess formal safe tracking guarantees. In principle, reachability analysis tools (such as those referenced in the Introduction) could be employed to verify its performance. However, the scalability of these methods is limited to low-dimensional systems and short time horizons, making them impractical for tracking applications, which typically involve a relatively large number of time steps. Therefore, in this work, we focus on formal safety verification of the NN-based tracking controller and provide statistical safety guarantees through conformal prediction.
\end{remark}

\subsection{Formal Verification with Conformal Prediction}\label{sec:FormalVerification}
To obtain rigorous guarantees on tracking performance under disturbances, we utilize \emph{formal verification} based on \emph{conformal prediction}, which is a nonparametric framework to provide distribution-free confidence sets for general machine learning models \cite{vovk2005algorithmic,shafer2008tutorial,angelopoulos2023conformal}.  Conformal prediction offers a unified framework for designing safe controllers and performing offline verification of learning-enabled autonomous systems, while maintaining scalability, interpretability, and formal guarantees \cite{lindemann2024formal}. This approach is especially well-suited for safety-critical systems, as it offers finite-sample coverage guarantees under mild assumptions.

We apply the standard full conformal prediction procedure to verify whether closed-loop trajectories satisfy the reach-avoid specification with a desired confidence level. Specifically, we first construct forward safe sets  $\mathcal{S}_k\subseteq\mathcal{X}_{\mathrm{o}}\setminus \mathcal{X}_{\mathrm{u}},~k\in \intcc{0;N-1},$ as follows.
For each time step $k \in \intcc{0;N-1}$, we define $\mathcal{S}_k \subset \mathbb{R}^n$ as an axis-aligned hyper-rectangle, containing the nominal state $\tilde{x}_k$ and ensuring that all states within $\mathcal{S}_k$ remain in the operating region $\mathcal{X}_{\mathrm{o}}$ and avoid the unsafe set $\mathcal{X}_{\mathrm{u}}$. We parameterize $\mathcal{S}_k$ by one-sided half-widths vectors $r_k^+, r_k^- \in \mathbb{R}_+^n$ representing the maximal allowable displacements from the nominal state $\tilde{x}_k$ for each coordinate in the positive and negative directions respectively. 
Each axis $j \in \{1,\dots,n\}$ is categorized to  either be a separating coordinate or a non-separating coordinate. The corresponding $(r_k^+)_j, (r_k^-)_j \in \mathbb{R}_+$ are then determined by the distance between the nominal coordinate $(\tilde{x}_k)_j$ and the nearest boundaries of $\mathcal{X}_{\mathrm{o}}$ and $\mathcal{X}_{\mathrm{u}}$. Specifically, \begin{equation}\label{eq:safe_hyperrect}
    \mathcal{S}_k \triangleq 
    \Hintcc{\tilde{x}_k - r_k^-,\, \tilde{x}_k + r_k^+},
\end{equation}
For separating coordinate $j$, we set
\begin{equation}
\begin{aligned} \label{eq:rk+,rk-,sep}
(r_k^+)_j &= \min\left\{
(\overline{x}_{\mathrm{o}})_j - (\tilde{x}_k)_j,
\min_{i \in \mathcal{A}^+_{k,j}} \left\{{(\underline{x}_{\mathrm{u}}^{(i)})_j - (\tilde{x}_k)_j}\right\}
\right\}- \varepsilon,\\
(r_k^-)_j &= \min\left\{
(\tilde{x}_k)_j - (\underline{x}_{\mathrm{o}})_j,
\min_{i \in \mathcal{A}^-_{k,j}} \left\{{(\tilde{x}_k)_j - (\overline{x}_{\mathrm{u}}^{(i)})_j}\right\}
\right\}- \varepsilon,
\end{aligned}
\end{equation}
where the active index sets are defined as $\mathcal{A}^+_{k,j}\triangleq\{ i:{(\underline{x}_{\mathrm{u}}^{(i)})_j - (\tilde{x}_k)_j}>0\}$ and 
$\mathcal{A}^-_{k,j}\triangleq \{ i:{(\tilde{x}_k)_j - (\overline{x}_{\mathrm{u}}^{(i)})_j}>0\}$.\\
For non-separating coordinate $j$, we set
\begin{equation}\label{eq:rk+,rk-,nonsep}
\begin{aligned}
    (r_k^+)_j &= (\overline{x}_{\mathrm{o}})_j - (\tilde{x}_k)_j - \varepsilon\\
    (r_k^-)_j &= (\tilde{x}_k)_j - (\underline{x}_{\mathrm{o}})_j - \varepsilon,
\end{aligned}
\end{equation}
Here, $\varepsilon>0$ is a small safety margin to maintain strict separation from obstacle boundaries. 
In practice, we select separating coordinates and non-separating coordinates such that $(r_k^+)_j\geq0,(r_k^-)_j\geq0$, $\forall k\in \intcc{0;N-1}, \forall j\in \intcc{1;n}$. Then, $\mathcal{S}_k$ is guaranteed to be strictly separated from $\mathcal{X}_{\mathrm{o}}\setminus \mathcal{X}_{\mathrm{u}}$.
\begin{lemma}\label{lem:separable}
For any given $k \in \intcc{0;N-1}$, suppose there exists at least one coordinate 
$j^s \in \{1,\dots,n\}$ such that the corresponding  $\mathcal{S}_k$ 
and every obstacle $\mathcal{X}_{\mathrm{u}}^{(i)}$ are strictly disjoint, i.e., for $\forall i \in \intcc{1;N_{\mathrm{u}}},$
\begin{equation}
    \Big[
        (\tilde{x}_k)_{j^s} - (r_k^-)_{j^s},\,
        (\tilde{x}_k)_{j^s} + (r_k^+)_{j^s}
    \Big]
    \cap
    \Big[
        (\underline{x}_{\mathrm{u}}^{(i)})_{j^s},\,
        (\overline{x}_{\mathrm{u}}^{(i)})_{j^s}
    \Big]
    = \varnothing.
\end{equation}
If, in addition, the one-sided half-widths satisfy $(r_k^+)_j \ge 0$ and $(r_k^-)_j \ge 0$ 
for all $j \in \intcc{1;n}$, then it is guaranteed 
\begin{equation}
    \mathcal{S}_k \subseteq \mathcal{X}_{\mathrm{o}}, 
    \mathcal{S}_k \cap \mathcal{X}_{\mathrm{u}}^{(i)} = \varnothing,
    \quad \forall i \in \intcc{1;N_{\mathrm{u}}}.
\end{equation}
\end{lemma}
\begin{remark}
    By Assumption  \ref{assm: XiXt}, the nominal trajectories satisfy $\tilde{x}_k\in\mathcal{X}_{\mathrm{o}}\setminus \mathcal{X}_{\mathrm{u}}$, $\forall i\in\intcc{1;N_{\mathrm{u}}}$, $k\in\intcc{0;N}$. Therefore, the conditions in Lemma \ref{lem:separable} are naturally satisfied.
\end{remark}

The resulting sets act as trajectory-level safety certificates in the sense that any trajectory $ \tau = (x_0, x_1, \dots, x_N)$ is guaranteed to satisfy the reach-avoid specification if
$$
x_k \in \mathcal{S}_k \text{ for all } k \in \intcc{0;N-1}, \quad \text{and} \quad x_N \in \mathcal{X}_{\mathrm{t}}.
$$
Then, to numerically encode specification violations, we define the stepwise violation at time step  $k$ as
\begin{equation}\label{eq:nonconformity_score}
s_k \triangleq
\begin{cases}
    \mathrm{dist}(x_k, \mathcal{S}_k) & \text{if } k \in \intcc{0;N-1}, \\
    \mathrm{dist}(x_N,\mathcal{X}_{\mathrm{t}})   & \text{if } k = N,
\end{cases}
\end{equation}
where where 
$\mathrm{dist}(x,\, \mathcal{A}) \triangleq \inf_{y \in \mathcal{A}} \|x - y\|_2$. We then have the \textit{trajectory-level nonconformity score} defined as the accumulated violation
\begin{equation} \label{eq:nonconformity_score total}
s(\tau) \triangleq \sum_{k=0}^N s_k.
\end{equation}
Consequently, $ s(\tau) = 0 $ if and only if the trajectory $\tau$ passes the safety certificate over the horizon. Larger values of $ s(\tau) $ indicate greater overall violation.

Now, let $ \{\tau_h\}_{h=1}^{H_0^{test}} $ be a test set of closed-loop trajectories, each generated by applying the tracking controller to the system~\eqref{eq:NonlinearSystem} with random initial condition $ x_{0,h} \in \Lambda_0 $ and disturbance sequence $ \{w_{k,h}\}_{k=1}^{N} \subseteq  \mathcal{W}^N $. Each trajectory $ \tau_h = (x_{0,h}, x_{1,h}, \dots, x_{N,h}) $ yields a score $s(\tau_h)$.
\begin{assumption}\label{assm:exchangeability}
The initial conditions $x_0^{(i)} \sim \Lambda_0$ are drawn independently, and the disturbances $w_k^{(i)} \sim \mathcal{W}$ are independent across time and rollouts. 
\end{assumption}

Under this  scheme, the resulting trajectories $\tau_i$ are independent thus exchangeable random elements \cite{bernardo1996concept} in the trajectory space, satisfying the conditions to apply conformal prediction to this problem.

Then, we compute a quantile threshold to certify any unseen (future) trajectory. Let the scores $\{s(\tau_h)\}_{h=1}^{H_0^{\text{test}}}$ be sorted in ascending order as $s_{(1)} \le \cdots \le s_{(H_0^{\text{test}})}$. For a confidence level $1 - \delta$, where $\delta$ is user-defined, we define the empirical quantile
$q_{1-\delta} \triangleq s_{(l)}$ with $l = \lceil (1 - \delta)(H_0^{\text{test}} + 1) \rceil.$ Denote $\tau_{\text{new}}$ to be a future rollout with initial state and disturbance $(x_{0,\text{new}}, \{w_{k,\text{new}}\}_{k=0}^{N-1})$ and associated score to be $s(\tau_{\text{new}})$. We have the result below.
\begin{proposition}[Reach-Avoid Guarantee~\cite{lindemann2024formal}]
\label{thm:reach-avoid-guarantee}
Under Assumption~\ref{assm:exchangeability}, the trajectory $\tau_{\text{new}}$ satisfies
\begin{equation}
\mathbb{P} \left[ s(\tau_{\text{new}}) \le q_{1-\delta} \right] \ge 1 - \delta.
\end{equation}
\end{proposition}

This provides a certified formal statistical safety guarantee for the tracking controller, purely from sampled trajectories, without requiring knowledge of the distribution over initial conditions or disturbances.

\section{Experiments} \label{sec:NumericalExamples}
Consider the discrete-time Dubin's car system:
{\small\begin{equation}
    \begin{split}
    \nonumber
     \left(   \begin{array}{c}
             (x_{k+1})_1 \\
             (x_{k+1})_2
             \\ 
             (x_{k+1})_3
        \end{array}\right)\in\left(   \begin{array}{c}
             (x_k)_1 +\tau_{s}(u_k)_1\cos((x_k)_3) \\
             (x_k)_2+\tau_{s}(u_k)_1\sin((x_k)_3)
             \\ 
             (x_k)_3+\tau_{s}(u_k)_2
    \end{array}\right)+\mathcal{W},
    \end{split}
\end{equation}}
where $((x_k)_1,(x_k)_2)$ denotes the car position, $(x_k)_3$ denotes the heading angle, $\tau_{s}$ is the sampling time ($\tau_{s}=0.05$),  $(u_{k})_{1}$ and $(u_{k})_{2}$ are control inputs corresponding to the car speed and turning rate, respectively, and the disturbance set is given by $\mathcal{W}=\tau_{s}(\intcc{-0.02,0.02}^{2}\times \intcc{-0.1,0.1})$. We now demonstrate how our approach can be used to synthesize a NN tracking controller that keeps the system states in the operating domain $\mathcal{X}_{\mathrm{o}}=\intcc{0,5}\times \intcc{0,2} \times   \intcc{-\pi/2,\pi/2}$, while avoiding the unsafe set $((\intcc{1.5,3}\times \intcc{0,0.5})\cup (\intcc{1.5,2.5}\times \intcc{1,2})\cup (\intcc{3.5,4.5}\times \intcc{0,1}))\times \intcc{-\pi/2,\pi/2}$, and steering the system's states to the target set $\mathcal{X}_{\mathrm{t}}=\intcc{3.5,5}\times \intcc{1.5,2} \times   \intcc{-\pi/5,\pi/5}$, using  control values in the set $\mathcal{U}=\intcc{-8,8}\times \intcc{-5,5}$. The initial set $\mathcal{X}_{\mathrm{i}}$ is singleton, corresponding to a starting point, that is assigned  different values discussed shortly.

For each starting point value, we calculate a nominal trajectory, using a sampling-based approach (RRT), and  obtain under-approximate BRSs along the nominal trajectory. We carry out the zonotopic computations, the zonotopic plots, and the sampling from zonotopes using the reachability toolbox CORA \cite{Althoff15CORA}.

The NN controller $\mu$ is trained as follows. At time step $k$, we generate $H_k^u = 40$ points for the uniform sampling set $\mathbf{S}_k^{u}$ and $H_k^{ex}=60$ points for the extreme sampling set $\mathbf{S}_k^{ex}$ using command \textit{randPoint}\cite{Althoff15CORA}. The entire sampling set $\mathbf{S}_k=\mathbf{S}_k^u\bigcup\mathbf{S}_k^{ex}$ is then randomly split into training set $\mathbf{S}_{k, train}$ and validation set $\mathbf{S}_{k, val}$ at a ratio of 7:3, with $H_k^{train}=70$ and $H_k^{val}=30$. During the training, validation is conducted every 200 epochs, and the training is terminated if there is no further improvement on the loss for 5 consecutive  validations, with the validation loss being 
$\frac{1}{H_k^{val}}\sum\limits_{z_{k,h} \in \mathbf{S}_{k, val}}\|d_{k,G}(z_{k,h})\|_{\infty}$
for time step $k$. 
 In the objective function, we set $\lambda=0.1$, $\alpha_{k,1}=10$, and $\alpha_{k,2}=0.01$.

For the Dubin's car, $n=3$ and $m=2$. Correspondingly, we set the number of neurons for each layer to be $d_0=d_5=3$, $d_1=d_2=d_3=d_4=32$, and $d_6=2$. The activation function $\phi$ in the first $l-2$ layers is ReLU. We apply Kaiming initialization \cite{he2015delving} for the weights and zero initialization for the biases. The training is performed using the Adam optimizer with a learning rate of $3\times10^{-4}$ and a weight decay of $10^{-4}$ on a local machine equipped with a 13th Gen Intel\textsuperscript{\textregistered} Core\textsuperscript{\texttrademark} i7-1360P processor featuring 12 parallel computing cores\footnote{The code for all experiments is located in the Github repository: https://github.com/YuezhuXu/Learning-NN-Safe-Tracking-Controllers-from-Backward-Reachable-Sets}. For formal verification, we set the safety margin in \eqref{eq:rk+,rk-,sep}, \eqref{eq:rk+,rk-,nonsep} to be $\varepsilon=0.01$ and perform the analysis under the same experimental setting.

To demonstrate the consistent performance of our NN controllers, we conduct experiments on three cases with different starting points  $[4,0.5,-\pi/4]^\top$, $[0.5,4.5,\pi/4]^\top$, and $[1,1,0]^\top$ respectively. 
For the three cases, we report the training time to be  8 minutes 35 seconds, 8 minutes 33 seconds, and 7 minutes 11 seconds respectively.

\subsection{Consistent and Reliable Performance of Online Neural Network Tracking Controllers}\label{sec:exp,consistent, reliable}
\begin{figure}[!htbp]
    \centering
    \begin{minipage}{0.42\textwidth}
        \includegraphics[width=\textwidth]{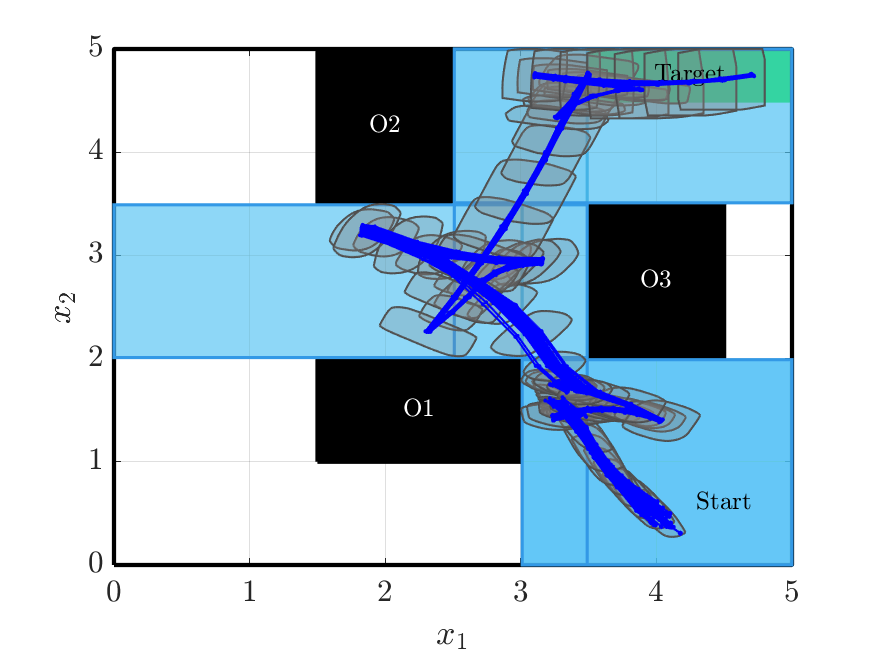}
    \end{minipage}
    \hfill
    \begin{minipage}{0.42\textwidth}
        \includegraphics[width=\textwidth]{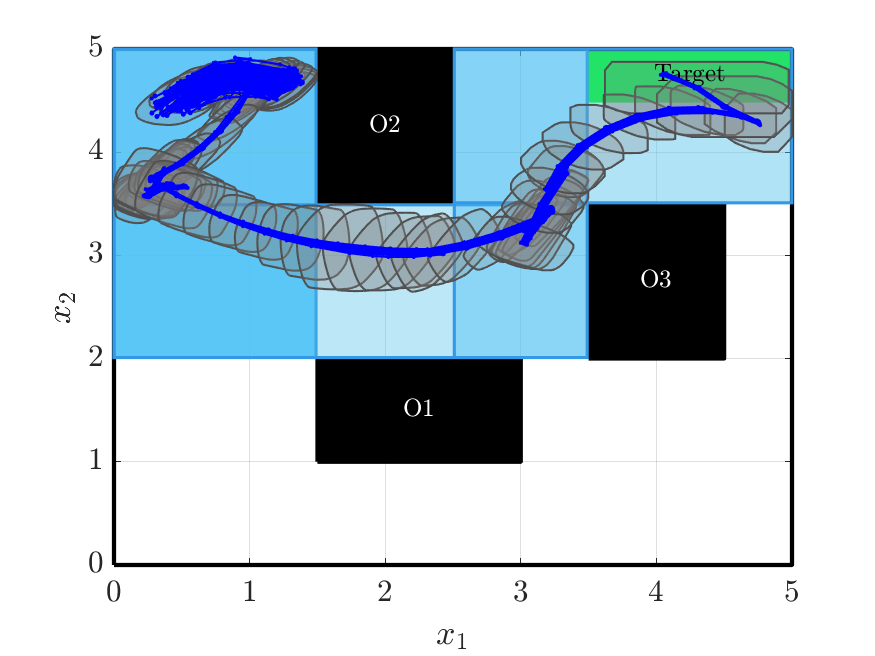}
    \end{minipage}
    \begin{minipage}{0.42\textwidth}
        \includegraphics[width=\textwidth]{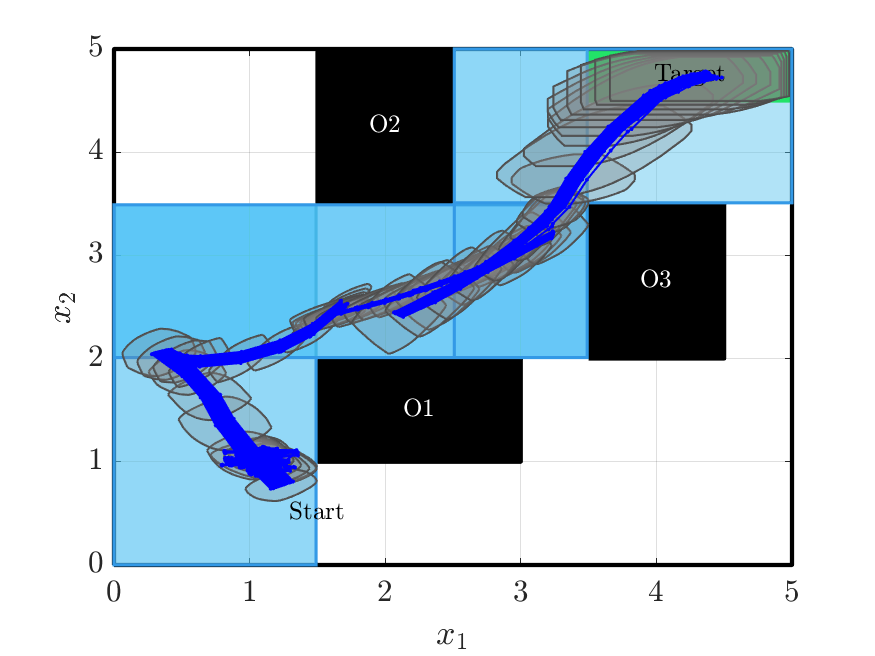}
    \end{minipage}
    \caption{$O_1,O_2,O_3$ are obstacles to avoid and the green set is the target set.
        The nominal starting points $\tilde{x}_0$ are:\\
        $[4,0.5,-\pi/4]^\top$, $[0.5,4.5,-\pi/4]^\top$, and $[1,1,0]^\top$ respectively. For visual clarity, only 100 trajectories are shown out of 1000 simulated.}
    \label{fig:regular uniform sampling}
    \vspace{-1em}
\end{figure}

With the trained NN controllers, we have a closed-form mapping from the states to the inputs. In other words, regardless of the starting points within $\Lambda_{0}$, we eliminate the need to solve any optimization problem during  execution, thus significantly enhancing computational efficiency for online control applications. 
To evaluate the performance of the trained controllers, we apply the full conformal prediction method described in Section \ref{sec:FormalVerification} for all three cases. As outlined in \eqref{eq:safe_hyperrect}-\eqref{eq:rk+,rk-,nonsep}, we set the first and second dimension to be the separating coordinates and the third dimension to be the non-separating coordinate. We then construct forward safe sets $\{\mathcal{S}_k\}_{k=0}^{N}$ as axis-aligned hyper-rectangles centered at the nominal states $\{\tilde{x}_k\}_{k=0}^{N-1}$, guaranteed to lie within the operating domain $\mathcal{X}_{\mathrm{o}}$ and avoid the unsafe set $\mathcal{X}_{\mathrm{u}}$ (the series of light blue boxes in Fig. \ref{fig:regular uniform sampling}). These sets act as safety certificates for each step. For each case, we simulate $H_0^{test} = 1000$ closed-loop trajectories $\{\tau_h\}_{h=1}^{H_0^{test}}$ under the fixed NN controller. The initial states $\{x_{0,h}\}_{h=1}^{H_0^{test}}$ are uniformly sampled from the corresponding initial zonotope $\Lambda_0 = \tilde{x}_0 + G^0 \mathbb{B}_{q_0}$, using the \textit{randPoint} command with the sampling type set to \texttt{'uniform'}. The disturbance is drawn independently at each step following the uniform distribution $\mathcal{W}=\mathbf{U}(-a,a)$, where $a = \begin{bmatrix}
    0.003, 0.003,0.015
\end{bmatrix}^T
$. For each case, we check that all $1000$ simulated trajectories stay inside $\mathcal{S}_k$ at stage $k$,  $\forall k\in\intcc{0;N-1}$ and reach $\mathcal{X}_t$ at last, i.e., $s_k=0$, $\forall k\in\intcc{0;N}$. According to \eqref{eq:nonconformity_score}-\eqref{eq:nonconformity_score total}, for our experiments, we obtain,
\begin{equation}
    s(\tau_h) = 0, \quad \forall h \in \intcc{1;1000}.
\end{equation}
As a result, the sorted scores are $s_{(1)} = \cdots = s_{(1000)} = 0$. To construct a conformal certification rule with confidence level $99.9\%$, $\delta = 0.001$, we compute
\begin{equation}
    l = \lceil (1 - \delta)(H_0^{test} + 1) \rceil = \lceil 0.999 \cdot 1001 \rceil = 1000,
\end{equation}
meaning that $q_{0.999} = s_{(1000)} = 0.$
Therefore, according to Proposition \ref{thm:reach-avoid-guarantee}, a future rollout $\tau_{\text{new}}$ has $s(\tau_{\text{new}}) \le q_{0.999}  = 0.$ In other words, any $\tau_{\text{new}}$  is certified to stay inside the forward safe sets $\mathcal{S}_k$ at stage $k$,  $\forall k\in\intcc{0;N-1}$ and reach $\mathcal{X}_t$ at last, thus satisfying the reach-avoid specification, at $99.9\%$ confidence.
In fact, owing to the learning ability of the NN controller, we manage to guide the states towards the center of zonotope for the next time step, resulting in different trajectories contracting to the nominal trajectory as the system evolves. Meanwhile, the simulation time for each trajectory is approximately 0.0270 seconds, 0.0245 seconds, and 0.0220 seconds respectively for the three cases. Note that if controls need to be generated for multiple initial states, there is potential for even faster implementation by leveraging parallel computing.

\begin{figure*}[t]
  \centering
  \begin{subfigure}[t]{0.3\textwidth}
    \centering
    \includegraphics[width=\textwidth]{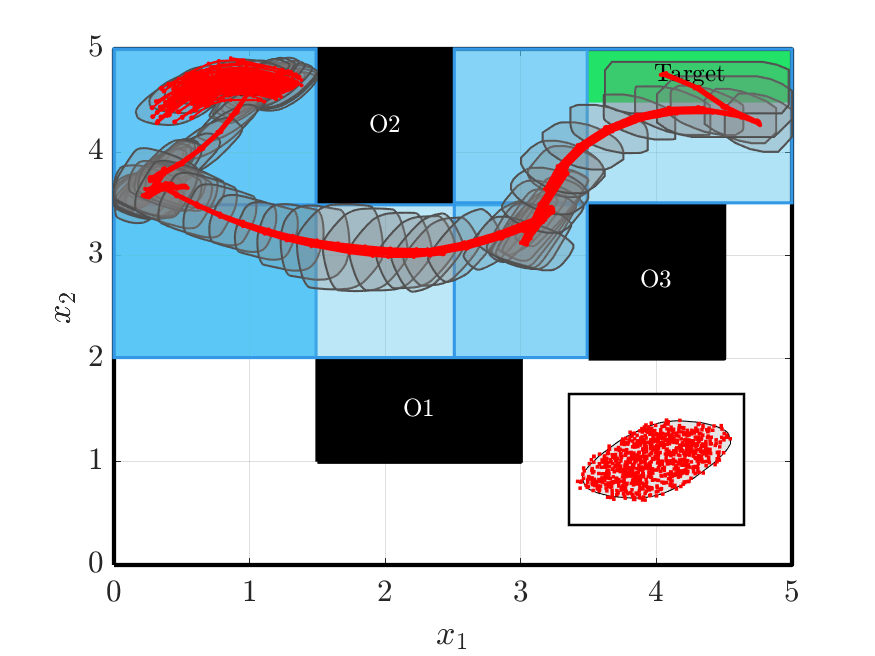}
    \caption{$\tau=0.1$}
    \label{fig:tau0.1}
  \end{subfigure}
  \hfill
  \begin{subfigure}[t]{0.3\textwidth}
    \centering
    \includegraphics[width=\textwidth]{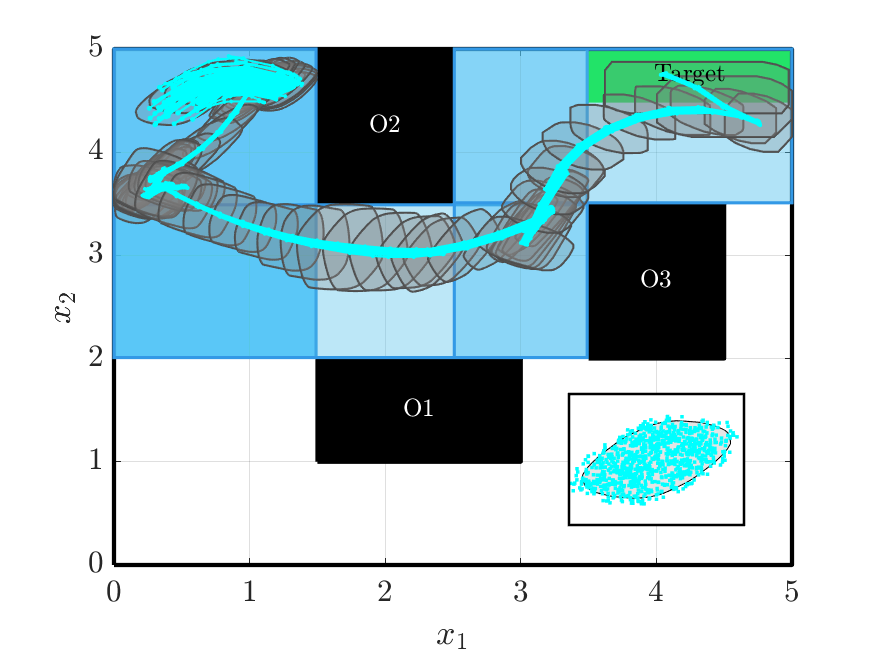}
    \caption{$\tau=0.2$}
    \label{fig:tau0.2}
  \end{subfigure}
  \hfill
  \begin{subfigure}[t]{0.3\textwidth}
    \centering
    \includegraphics[width=\textwidth]{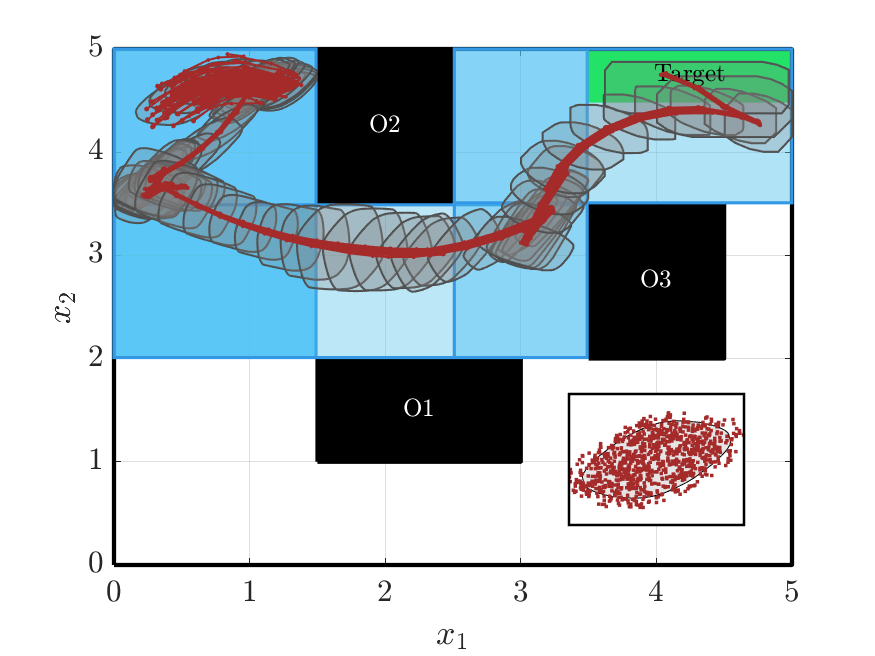}
    \caption{$\tau=0.3$}
    \label{fig:tau0.3}
  \end{subfigure}
  \caption{Trajectories generated with the initial states uniformly sampled from the enlarged initial zonotopes $\Lambda_0^{\tau}$. For visual clarity, only 100 trajectories are shown out of 1000 simulated for each case. The original initial zonotope and the sampled initial states are highlighted in the inset in each plot.}
  \label{fig:comparison}
\end{figure*}

\begin{figure}[!htbp] 
    \centering
    % First plot
     \vspace{-1em}
    \includegraphics[width=0.45\textwidth]{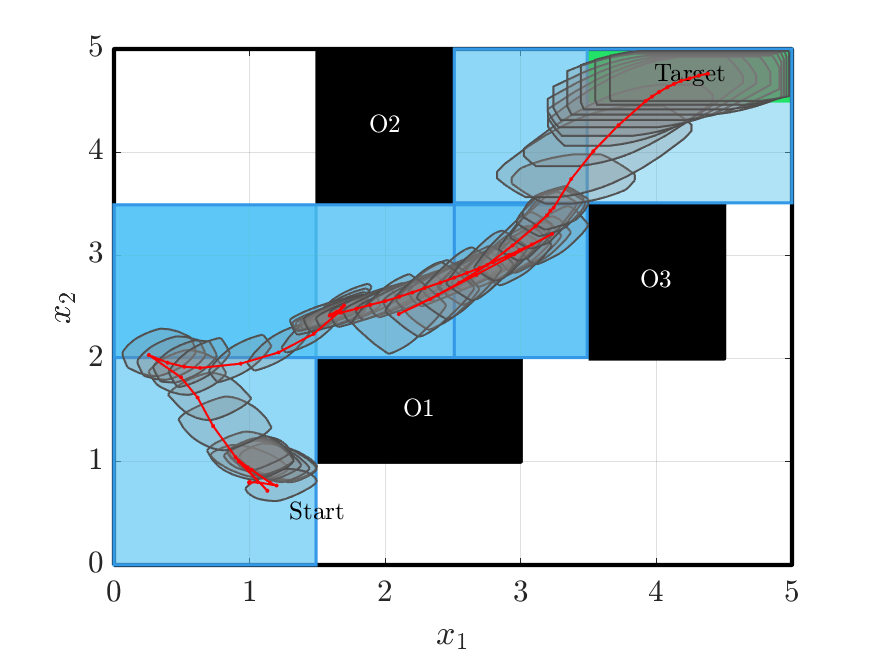} 
    \caption{$O_1,O_2,O_3$ are obstacles to avoid and the green set is the target set. The nominal starting points is $[1,1,0]^\top$ and the inital point we pick outside $\Lambda_0$ is $[1,0.8,0]^T$.}
    \label{fig:benchmark}
    \vspace{-1em}
\end{figure}

\subsection{Extending beyond Backward Reachable Sets}
Here, we demonstrate that the NN controller can have larger starting sets than optimization-based tracking controllers with BRSs. In other words, our NN controller is \textit{able to safely drive  points outside $\Lambda_0$ to the target set}. This highlights the strong generalization ability of the NN controllers in achieving superior tracking.
To illustrate this, we enlarge the scale of the initial zonotope $\Lambda_0=\tilde{x}_0+G^{0}\mathbb{B}_{q_{0}}$ for the case of the starting point $[0.5,4.5,-\pi/4]^\top$ by $10\%,20\%, 30\%,$ and $50\%$ respectively. The enlarged zonotopes are written as $\Lambda^\varsigma_0=\tilde{x}_0+(1+\varsigma)G^{0}\mathbb{B}_{q_{0}}$, with $\varsigma$ selected as $0.1,0.2$, and $0.3$. Then we sample from $\Lambda^\varsigma_0$ uniformly to obtain 1000 test points for each $\varsigma$. In Fig. \ref{fig:comparison}, we mark the trajectories for $\varsigma = 0.1,0.2,0.3$ with color red, and brown separately. We include a inset in each plot to highlight the initial zonotope and the distribution of starting points sampled.
Clearly, from  Fig. \ref{fig:comparison}, many initial points sampled are outside $\Lambda_0$ for $\varsigma=0.1,0.2$, and $0.3$. We observe that all $1000$ rollouts in each enlarged case $\varsigma$ remain inside the safe sets $\mathcal{S}_k$ (the series of light blue boxes same as the ones in Section \ref{sec:exp,consistent, reliable}) for $\forall k \in \intcc{0;N-1}$ and reach the target set $\mathcal{X}_t$, meaning that each trajectory still achieves $s^\varsigma_k = 0$ for all $k$, and thus
\[
s^\varsigma(\tau_h) = 0, \quad \forall h \in \intcc{1;1000}.
\]
This again implies that the sorted scores satisfy $s^\varsigma_{(1)} = \cdots = s^\varsigma_{(1000)} = 0$. For each $\varsigma$, we set $\delta = 0.001$ to construct a conformal predictor at $99.9\%$ confidence, and compute the quantile
\[
l = \lceil (1 - \delta)(n + 1) \rceil = \lceil 0.999 \cdot 1001 \rceil = 1000,
\]
with $q_{0.999} = s_{(1000)} = 0$. Therefore, for each enlarged initial set $\Lambda_0^\varsigma$, any future rollout $\tau^\varsigma_{\text{new}}$  is certified to satisfy $s^\varsigma(\tau^\varsigma_{\text{new}}) \le 0$, thus meeting the reach-avoid specification, with $99.9\%$ confidence. The simulation time for each trajectory remains low, being approximately 0.0252 seconds, 0.0258 seconds,  and 0.0263 seconds, respectively for $\tau=0.1,0.2$, and $0.3$.

Next, we benchmark the performance of the NN controllers against the optimization-based controller from \cite{serry2024safe}.
We consider the case of the starting nominal point $[1,1,0]^\top$ and pick an initial point $[1,0.8,0]^\top$ outside $\Lambda_{0}$. The optimization-based controller from \cite{serry2024safe} fails in the sense that it cannot find a control value that drives the initial value to $\Lambda_{1}$. Starting from the same position, we apply trained NN tracking controller which manages to safely drive the state to the target set, even when the starting point is outside $\Lambda_{0}$ as shown in Fig. \ref{fig:benchmark}.

This provides empirical evidence suggesting that with the use of NN controllers,  safe sets can be  expanded well beyond the theoretical guarantees provided by classical backward reachability based control designs, motivating the need for further research in closing the gap between provable guarantees and the tracking performance observed in practice.

\section{Discussion and Future Work}\label{sec:Conclusion}
In this work, we presented a NN-based safe tracking control framework for nonlinear discrete-time systems with  reach-avoid specifications in the presence of disturbances. Our approach relies on nominal  trajectory generation  and computations of safe zonotopic BRSs along the nominal trajectory. The computed BRSs are then used to guide the training of a NN tracking controller, such that the NN drives the system's states through these BRSs, thus increasing the likelihood of safe reachability. We also provide statistical guarantees on the safe tracking capabilities of the learned NN controller with a high confidence level of $99.9\%$.  
Motivated by this work, we aim to explore several research directions in the future. One of the direction is refining the computation of BRSs such as those in \cite{serry2024safe} so that they can better aid NN tracking control by providing larger sets of training points. We will also explore new BRS-informed NN architecture designs that can potentially provide better tracking capabilities.

\balance

\bibliographystyle{elsarticle-num}
%\bibliography{references} 

\end{document}